\documentstyle[12pt]{article}

\begin{document}
\begin{center}
{\large\bf Cosmological Constant Problem}
\vskip 0.3 true in
{\large J. W. Moffat}
\vskip 0.3 true in
{\large Perimeter Institute for Theoretical Physics, Waterloo,
Canada} \vskip 0.3 true in and \vskip 0.3 true in {\large
Department of Physics, University of Toronto, Ontario, Canada}
\end{center}

\begin{abstract}
The cosmological constant problem is reviewed and a possible
quantum gravity resolution is proposed. A space satellite
E\"otv\"os experiment for zero-point vacuum energy is proposed to
see whether Casimir vacuum energy falls in a gravitational field
at the same rate as ordinary matter. \end{abstract}

 \section{\bf Cosmological Constant Problem}

$\bullet$ It is generally agreed that the cosmological constant
problem (CCP) is one of the most severe problems facing modern
particle and gravitational physics. It is believed that its
solution could significantly alter our understanding of particle
physics and cosmology~\cite{Straumann}.

$\bullet$ The accelerating universe and dark energy. Is the
cosmological constant the explanation for the accelerating
universe? What is dark energy?

$\bullet$  There have been many attempts to solve the CCP.
Adjustment models do not avoid fine-tuning.

$\bullet$  Higher-dimensional models of the brane-bulk type do not avoid
fine-tuning.

$\bullet$ Superstring theory has not yet provided a solution to
the CCP.

In the following, I will describe a possible resolution of
the CCP, based on a model of a nonlocal
field theory and quantum gravity theory that suppresses the coupling of
gravity to vacuum energy density. The violation of the equivalence
principle in the theory can be tested by performing E\"otvos experiments on
Casimir vacuum energy in satellites.

\section{\bf Gravitational Coupling to Vacuum Energy and Quantum
Gravity Theory}

We can define an effective cosmological constant
\begin{equation}
\lambda_{\rm eff}=\lambda_0+\lambda_{\rm vac},
\end{equation}
where $\lambda_0$ is the ``bare'' cosmological
constant in Einstein's classical field equations,
and $\lambda_{\rm vac}$ is the contribution that arises from the
vacuum density $\lambda_{\rm vac}=8\pi G\rho_{\rm vac}$.

Already at the standard model (SM) electroweak scale $\sim 10^2$
GeV, a calculation of the vacuum density $\rho_{\rm vac}$, based
on local quantum field theory results in a discrepancy of order
$10^{55}$ with the observational bound $\rho_{\rm vac} \leq
10^{-47}\, ({\rm GeV})^4 \sim (10^{-3}\,eV)^4$. The WMAP and
supernovae SNIa data require dark energy~\cite{Spergel}. If the
vacuum energy is the dark energy, then $\rho_{\rm vac}\sim
(10^{-3}\, eV)^4$. {\it There is an egregious discrepancy
between the particle physics estimate of $\rho_{\rm vac}$ and
the cosmological observation}.

There is a severe fine-tuning problem of order $10^{55}$, since
the virtual quantum fluctuations giving rise to $\lambda_{\rm vac}$ must
cancel $\lambda_0$ to an unbelievable degree of accuracy.
This is the ``particle physics'' source of the cosmological
constant problem.

\section{\bf Nonlocal Quantum Gravity Model}

Let us consider a model of nonlocal gravity with the
action~\cite{Moffat,Gillies}:
\begin{equation}
S=S_G+S_M,
\end{equation}
where ($\kappa^2=8\pi G$):
\begin{equation}
S_G=-\frac{1}{2\kappa^2}\int d^4x\sqrt{-g}(R[g,{\cal
G}^{-1}]+{\cal G}^{-1}2\lambda_0)
\end{equation}
and
\begin{equation}
S_M=\frac{1}{2}\int d^4x\sqrt{-g}{\cal
G}^{-1}\biggl(g^{\mu\nu}\nabla_\mu\phi{\cal F}^{-1}\nabla_\nu\phi
-m^2\phi{\cal F}^{-1}\phi\biggr).
\end{equation}

${\cal G}$ and ${\cal F}$ are nonlocal regularizing,
{\it entire} functions. As an example, we can choose the covariant
functions
\begin{equation}
{\cal G}(x)=\exp\biggl[-{\cal D}(x)/2\Lambda_G^2\biggr],
$$ $$
{\cal F}(x)=\exp\biggl[-({\cal D}(x)+m^2)/2\Lambda_M^2)\biggr],
\end{equation}
where ${\cal D}\equiv\nabla_\mu\nabla^\mu$ and $\Lambda_G$ and $\Lambda_M$
are $({\rm length})^{-1}$ (energy) scales.

We expand $g_{\mu\nu}$ about flat Minkowski spacetime:
$g_{\mu\nu}=\eta_{\mu\nu}+2\kappa h_{\mu\nu}$. The propagators for the
graviton and the $\phi$ field in a fixed gauge are given by
\begin{equation}
D^\phi(p)=\frac{{\cal G}(p){\cal F}(p)}{p^2-m^2+i\epsilon},
\end{equation}
$$ $$
\begin{equation}
D^G_{\mu\nu\rho\sigma}(p)=\frac{(\eta_{\mu\rho}\eta_{\nu\sigma}
+\eta_{\mu\sigma}\eta_{\nu\rho}-\eta_{\mu\nu}\eta_{\rho\sigma})
{\cal G}(p)}{p^2+i\epsilon}.
\end{equation}

Because ${\cal G}$ and
${\cal F}$ are {\it entire} functions of $p^2$, preserving the Cutkosky
rules, they do not violate unitarity.
Gauge invariance can be maintained by satisfying certain
constraint equations for ${\cal G}$ and ${\cal F}$ in every order of
perturbation theory.  This guarantees that $\nabla_\nu T^{\mu\nu}=0$.

\section{\bf Resolution of the Cosmological Constant Problem}

In flat Minkowski spacetime, the sum of all {\it disconnected}
vacuum diagrams $C=\sum_nM^{(0)}_n$ is a constant factor in the
scattering S-matrix $S'=SC$. Since the S-matrix is unitary
$\vert S'\vert^2=1$, then we must conclude that $\vert
C\vert^2=1$, and all the disconnected vacuum graphs can be
ignored. This result is also known to follow from the Wick ordering of the field
operators.

Due to the equivalence principle {\it gravity couples to all
forms of energy}, including the vacuum energy density $\rho_{\rm vac}$, so we can no
longer ignore these virtual quantum fluctuations in the presence of a non-zero
gravitational field.

Quantum corrections to $\lambda_0$ come from
loops formed from massive SM states, coupled to external
graviton lines at essentially zero momentum.

Consider the dominant contributions to the vacuum
density arising from the graviton-SM loop corrections.
We shall adopt a simple model consisting of a massive scalar
meson $\phi$, which has the SM mass $m\sim 10^2$ GeV.

The lowest order correction to the graviton-scalar vacuum loop
will have the form (in Euclidean momentum space):
\begin{equation}
\label{Ptensor}
\Pi^{\rm {\rm Gvac}}_{\mu\nu\rho\sigma}(p)=-\kappa^2
\int\frac{d^4q}{(q^2+m^2)[(q-p)^2+m^2]}
$$ $$
\times K_{\mu\nu\rho\sigma}(p,q)
\exp\biggl\{-(q^2+m^2)/2\Lambda^2_M
$$ $$
-[(q-p)^2+m^2]/2\Lambda^2_M-q^2/2\Lambda^2_{{\rm Gvac}}\biggr\}.
\end{equation}

For $\Lambda_{{\rm Gvac}} \ll \Lambda_M$, we observe that from
power counting of the momenta in the loop integral, we get
\begin{equation}
\Pi^{\rm {\rm Gvac}}_{\mu\nu\rho\sigma}(p)\sim
\kappa^2\Lambda_{{\rm Gvac}}^4N_{\mu\nu\rho\sigma}(p^2)
$$ $$
\sim\frac{\Lambda_{{\rm Gvac}}^4}{M^2_{\rm PL}}N_{\mu\nu\rho\sigma}(p^2),
\end{equation}
where $N(p^2)$ is a finite remaining part of $\Pi^{\rm {\rm Gvac}}(p)$.

We now have
\begin{equation}
\rho_{\rm vac}\sim M^2_{PL}\Pi^{\rm {\rm Gvac}}(p^2)\sim
\Lambda_{{\rm Gvac}}^4.
\end{equation}
If we choose $\Lambda_{{\rm Gvac}}\leq 10^{-3}$ eV, then the quantum correction
to the bare cosmological constant $\lambda_0$ is suppressed sufficiently
to satisfy the observational bound on $\lambda$, {\it and it is protected
from large unstable radiative corrections}.

This provides a solution to the
cosmological constant problem at the energy level of the SM
and possible higher energy extensions of the SM. The universal
fixed gravitational scale $\Lambda_{{\rm Gvac}}$ corresponds to the fundamental
length $\ell_{{\rm Gvac}}\leq 1$ mm at which virtual gravitational radiative
corrections to the vacuum energy are cut off.

The gravitational form factor ${\cal G}$, {\it when
coupled to non-vacuum SM gauge boson or matter loops}, will have the form
in Euclidean momentum space
\begin{equation} {\cal G}^{\rm GM}(q^2)
=\exp\biggl[-q^2/2\Lambda_{GM}^2\biggr].
\end{equation}

If we choose $\Lambda_{GM} = \Lambda_{M}> 1-10$ TeV, then we will
reproduce the SM experimental results, including the running
of the SM coupling constants, and ${\cal G}^{GM}(q^2)={\cal
F}^M(q^2)$ becomes ${\cal G}^{GM}(0)={\cal F}^M(q^2=m^2)=1$ on the mass
shell.

{\it This solution to the CCP leads to a violation of the weak equivalence
principle (WEP) for coupling of gravitons to vacuum energy and matter.} This could be
checked experimentally in a satellite E\"otvos experiment on the Casimir vacuum
energy.

We observe that the required suppression of the vacuum diagram
loop contribution to the cosmological constant, associated with
the vacuum energy momentum tensor at lowest order,
demands a low gravitational energy scale $\Lambda_{{\rm Gvac}}\leq
10^{-3}$ eV, which controls the coupling of gravitons to
pure vacuum graviton and matter fluctuation loops. This is
essentially because the external graviton momenta are close to the mass
shell, requiring a low energy scale $\Lambda_{{\rm Gvac}}$.

In our finite, perturbative
quantum gravity model nonlocal gravity produces a long-distance
infrared cut-off of the vacuum energy density through the low energy
scale $\Lambda_{{\rm Gvac}} < 10^{-3}$ eV~\cite{Sundrum}.
Gravitons coupled to {\it non-vacuum} matter tree graphs and
matter loops are controlled by the energy scale:
$\Lambda_{GM}=\Lambda_{SM} > 1-20$ TeV.

The rule is: When external graviton lines are removed from a
matter loop, leaving behind {\it pure} matter fluctuation vacuum loops,
then those initial graviton-vacuum loops are suppressed by the form factor
${\cal G}^{{\rm Gvac}}(q^2)$ where $q$ is the internal matter loop momentum and
${\cal G}^{{\rm Gvac}}(q^2)$ is controlled by $\Lambda_{{\rm Gvac}} \leq 10^{-3}$ eV.
On the other hand, e.g. the proton first-order self-energy graph, coupled
to a graviton, is controlled by $\Lambda_{GM}=\Lambda_M > 1-20$ TeV {\it
and does not lead to a detectable violation of the equivalence
principle.}

There are problematic issues associated with our nonlocal
quantum gravity model. Since the nonlocal form factors ${\cal G}_{\rm
Gvac}$ and ${\cal F}_M$ contain significantly different nonlocal energy
scales $\Lambda_G$ and $\Lambda_M$, unitarity at every order of
perturbation theory could pose problems. This requires further
study.

Complete S-matrix scattering amplitudes can through crossing
symmetry lead to large violations of causality for energies $\gg\Lambda$.
The nonlocal quantum gravity model should only be considered an effective
theory that regularizes the quantum gravity perturbation
calculations~\cite{Joglekar}.

Gluon condensates $\langle G^{\mu\nu}_a G_{a\mu\nu}\rangle_0$
formed in the QCD vacuum in phase transitions, due to a broken phase of
chiral symmetry, produce a vacuum density that is far too large: $\sim
\Lambda^4_{\rm QCD}/16\pi^2\sim 10^{-4}\, {\rm GeV}^4$, which is more than
40 orders of magnitude larger than $\rho_{\rm crit}$. The SM
Higgs particle produces a Higgs condensate $V(\phi=v)=-m^4/2\Lambda_c+V_0$
which is catastrophically large. These are both non-perturbative phenomena.
How do we explain a suppression of these condensates in a perturbative
quantum gravity scheme?

$\bullet$ The scales $\Lambda_M$ and $\Lambda_{{\rm Gvac}}$ are determined in
loop diagrams by the quantum non-localizable nature of the gravitons and
SM particles.

$\bullet$ The gravitons coupled to matter and matter
loops have a nonlocal scale at $\Lambda_{GM}=\Lambda_M > 1-20$ TeV or a
length scale $\ell_M < 10^{-16}$ cm, whereas the gravitons coupled
to pure vacuum energy are localizable up to an energy scale $\Lambda_{{\rm
Gvac}}\sim 10^{-3}$ eV or down to a length scale $\ell_{{\rm Gvac}} > 1$
mm.

$\bullet$ The fundamental energy scales $\Lambda_{{\rm Gvac}}$ and
$\Lambda_{GM}=\Lambda_M$ are determined by the underlying physical nature
of the particles and fields and do not correspond to arbitrary cut-offs,
which destroy the gauge invariance, Lorentz invariance and unitarity of the
quantum gravity theory for energies $>\Lambda_{{\rm Gvac}}\sim 10^{-3}$ eV.
The underlying explanation of these physical scales must be sought in a
more fundamental theory.

\section{\bf Satellite E\"otv\"os Experiment for Zero-point Vacuum
Energy}

We consider that the
cosmological constant arises from zero-point vacuum energy, so
that a violation of the WEP could be observed in an E\"otv\"os
experiment~\cite{Ross}. We propose~\cite{Gillies},
that a satellite experiment be performed in which the
acceleration of a spherical thin shell of aluminum be compared
to a test mass made of copper or silver. Aluminum has a sharp
transition from reflectance to absorption of EM waves at photon
energies of 15.5 eV. In a simple cutoff calculation, the
magnitude of the missing zero-point energy density inside the
aluminum sphere is
\begin{equation}
{\cal E}=\frac{4\pi}{(\hbar c)^3}\int_0^{{\rm E}_{\rm max}}dEE^3,
\end{equation}
where $E_{\rm max}$ is the energy at which aluminum becomes transparent.

For $E_{\rm max}=15.5$ eV, one obtains ${\cal E}=2.37\times
10^{19}\,eV/{\rm cm}^3$. The rest-mass energy density is $1.52\times
10^{33}\,eV/{\rm cm}^3$. Thus, the ratio is
\begin{equation}
R=1.6\times 10^{-14}.
\end{equation}
In drag-free satellite missions such as SEE, STEP, Galileo Galilei or
MICROSCOPE, the E\"otv\"os parameter $\eta=2(a_1-a_2)/(a_1+a_2)$
can reach an accuracy of $\eta\sim 10^{-15} - 10^{-17}$, so that our
Casimir vacuum energy E\"otv\"os test could reach a 1\% level.

The calculation of the Casimir vacuum energy for a thin spherical
shell of matter is controversial, due to the non-trivial self-energy
problem. For a thin hollow sphere a calculation of the Casimir
energy depends on the radius of the sphere in a non-trivial
manner and the sharp Dirichlet boundary conditions on the surface
of the sphere cause the calculation to be dependent on the
material of the sphere~\cite{Jaffe}. The calculation of the
zero-pont vacuum energy for a thin hollow sphere is cutoff
dependent, due to the emergence of divergences that cannot be
removed by a renormalization scheme.

$\bullet$ If the E\"otv\"os experiment shows that the vacuum energy
falls at a significantly slower rate than
ordinary matter in a gravitational field, then this is a strong indication
that the coupling of gravity to the vacuum energy is suppressed compared
to its coupling to ordinary matter. This would provide a significant clue
as to the basic mechanism that results in a small cosmological constant.

\section{\bf Conclusions}

$\bullet$ We have described a possible solution to the
cosmological constant problem. The particle physics resolution
requires that we construct a consistent quantum gravity theory,
which has vertex form factors that are different for gravitons
coupled to quantum {\it vacuum} fluctuations and matter.

$\bullet$ This predicts a violation of the WEP for
coupling to vacuum energy, but not to matter-graviton
couplings or to {\it non-vacuum matter loops}. This leads to a suppression
of all SM vacuum loop contributions and, thereby, avoids a
fine-tuning cancellation between the ``bare'' cosmological constant
$\lambda_0$ and the vacuum contribution $\lambda_{\rm vac}$. It retains the
experimental agreement of the SM predictions.

$\bullet$ A satellite
E\"otvos experiment for Casimir vacuum energy could experimentally decide
whether nature does allow a vacuum energy WEP violation and a significant
suppression of vacuum energy density.

$\bullet$ As a model of a future fundamental, nonlocal quantum gravity
theory, it does provide clues as to the resolution of the ``infamous''
cosmological constant problem.

{\bf Acknowledgments}

This work was supported by the Natural Sciences and Engineering
Research Council of Canada.


\begin{thebibliography}{100}

\bibitem{Straumann} S. Weinberg, Rev. Mod. Phys. {\bf 61}, 1
(1989); N. Straumann, astro-ph/020333.

\bibitem{Spergel} D. N. Spergel et al., astro-ph/0302209; S.
Perlmutter et al., Astrophys. J. {\bf 517}, 565 (1999); A. G.
Riess et al., Astron. J. {\bf 116}, 1009 (1998).

\bibitem{Moffat} J. W. Moffat, hep-ph/0102088; J. W. Moffat, AIP Conf.
Proc. {\bf 646}, 130 (2003), hep-th/0207198.

\bibitem{Gillies} J. W. Moffat and G. T.
Gillies, New J. Phys. {\bf 4} (2002) 92.1, gr-qc/0208005.

\bibitem{Sundrum} R. Sundrum, JHEP 9907 (1999) 001, hep-ph/9708329;
hep-th/0306106. In these papers, Sundrum proposes that the
graviton is a composite object with an associated form factor
and the cutoff $\Lambda_c \leq 10^{-3}$ eV plays a fundamental
role in suppressing the vacuum energy.

\bibitem{Joglekar} A. Jain and S. D. Joglekar, hep-th/0307208.

\bibitem{Ross} D. K. Ross, Nuovo. Cim. {\bf B114}, 1073 (1999).

\bibitem{Jaffe} R. Jaffe, Talk given at the QFEXT03 Workshop,
University of Oklahoma, Oklahoma, USA, September 15-19, 2003.
To be published in the workshop proceedings.

\end{thebibliography}
\end{document}